\documentclass[prl,twocolumn,showpacs]{revtex4}

\usepackage{amssymb,amsmath,bm,epsfig,graphicx,subfigure}

\begin{document}

\title{Stacking Textures and Singularities in Bilayer Graphene}
\author{Xingting Gong and E. J. Mele}
    \email{mele@physics.upenn.edu}
    \affiliation{Department of Physics and Astronomy University of Pennsylvania Philadelphia PA 19104  \\}
\date{\today}

\begin{abstract}
We study a family of globally smooth spatially varying two dimensional stacking textures in bilayer graphene. We find that the
strain-minimizing stacking patterns connecting inequivalent ground states with local $AB$ and $BA$ interlayer registries are
two dimensional twisted textures of an interlayer displacement field. We construct and display these topological stacking
textures for bilayer graphene, examine their interactions and develop the composition rules that allow us to represent other
more complex stacking textures, including globally twisted graphenes and extended one dimensional domain walls.
\end{abstract}

\pacs{61.48.Gh,61.72.Bb,61.65.Pq} \maketitle

\medskip
Bilayer graphene (BLG) features special functionalities that microscopically derive from various forms of broken sublattice
symmetry present when graphene sheets are stacked. These depend on relative lateral translations \cite{McCann,Ohta,MM},
rotations \cite{Lopes,meleRC} and layer symmetry breaking that can occur spontaneously \cite{layersymm,rahul,fanzhang2} or be
induced \cite{castro,zhang,Martinetal,zhangpnas,kim,Yacoby1,Yacoby2,Henriksen,pablo,Velasco,pablo2,Baoetal}. There has been
important recent progress imaging the stacking order in BLG using dark field transmission electron microscopy \cite{alden,lin}.
These experiments reveal rich submicron domain structures with locally registered $AB$ and $BA$ regions delineated by dense
irregular networks of domain walls, focusing attention on the inevitable competition between intralayer strain and interlayer
stacking commensuration energies \cite{alden,popov}.

In this Letter we examine a general family of  two dimensional stacking textures in BLG and their defects.  We find that the
strain minimizing stacking patterns that connect inequivalent ground states are twisted textures of the interlayer displacement
field.  We construct and display these topological stacking textures in BLG, examine their interactions and develop the
composition rules that allow us to represent other observed complex stacking textures, including twisted graphenes and extended
one dimensional domain walls.

A relative translation between two graphene layers is represented by an interlayer displacement vector $\Delta = f_1 T_1 +f_2
T_2$ where $T_{1,2} = a \exp(\pm i\pi/3)$ are primitive translation vectors of a graphene lattice with lattice constant $a$.
The lowest energy uniformly translated structures (Fig. 1) align the $A$ and $B$ sublattices of the two layers at $(f_1,f_2) =
(2/3,1/3)$ ($AB$ stacking: $\Delta_\alpha$) and its complement $(1/3,2/3)$ ($BA$ stacking: $\Delta_\beta$). The interlayer
potential is a periodic function $U(\Delta) = u_0 + u_1 \sum_n \, \exp\left(i \left( \bar G_n \Delta + G_n \bar \Delta
\right)/2\right)$ where $G_n = (4 \pi/\sqrt{3} a) \exp({i(2n-1)\pi/6})$ are vectors in the first star of reciprocal lattice
vectors. For BLG $u_1 \simeq 2.1 \, {\rm meV/atom}$ and $u_0 =3 u_1$ assigns the zero of energy to the $\alpha$- and
$\beta$-stacked states \cite{popov}.

\begin{figure}
  \includegraphics[angle=0,width=0.66\columnwidth,bb=0 0 1784 1800]{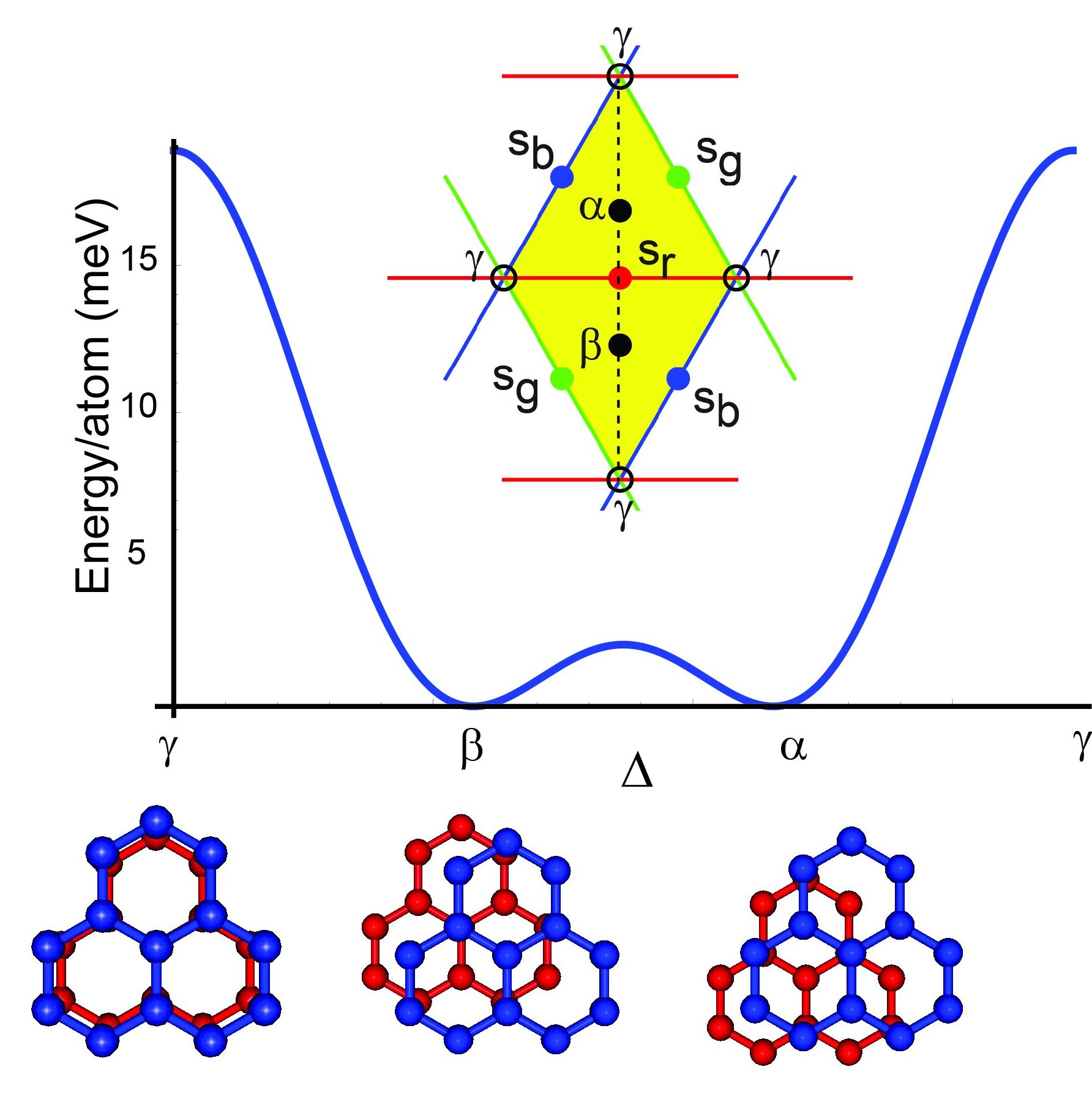}
  \caption{\label{geometry} A relative lateral interlayer translation is
  represented by a vector $\Delta=f_1 T_1 + f_2 T_2$  (with $T_{1(2)} = a e^{\pm i \pi/3}$) inside a fundamental domain (shaded rhombus) with the states at $\alpha, \, \beta, \, \gamma$ representing  $AB$, $BA$ and $AA$ stacking
  respectively. The lineplot gives the commensuration energy per atom on a vertical slice
  through connecting the three high symmetry states as shown. The points $s_r$, $s_g$ and $s_b$ are the saddle points
  on the potential energy surface.}
\end{figure}

The configuration space for $\Delta$ has the topology of a torus. When $u_1$ is large the energy minima at $\alpha$ and $\beta$
are deep and the shortest trajectories $\Delta(\vec r)$ connecting the two inequivalent minima $\alpha$ and $\beta$ are three
saddle point paths crossing the points labelled $s_r$, $s_g$ and $s_b$ in Fig. 1, indexed by their winding on the two cycles of
the torus. Each saddle point trajectory is bisected by a (straight) domain wall which runs along one of the three
symmetry-related directions as shown.

We focus on field textures $\Delta$ that satisfy the boundary conditions $\Delta(\infty) = \Delta_\alpha$ and $\Delta(0) =
\Delta_\beta$. A simple texture that accomplishes this wraps a stacking domain wall \cite{zhangpnas,kim} into a loop thereby
reversing the stacking order within a confined region, as shown in Fig. 2 for the field  $\Delta_w(z)=(1/2) ( \Delta_\alpha
+\Delta_\beta - (\Delta_\beta - \Delta_\alpha) \tanh [(|z|- R)/\ell])$ where $z=x+iy$ is the complex coordinate in the plane.
This texture connects two ground states through a transition region of width $\ell$  accumulating lattice strain in an annulus.
This texture passes through the value $(\Delta_\alpha + \Delta_\beta)/2$ on any radial path at $|z|=R$ and the wall evolves
smoothly from tensile character to shear character as a function of $ {\rm arg} [z]$ \cite{alden}.
\begin{figure}
  \includegraphics[angle=0,width=0.90\columnwidth,bb=0 0 1852 729]{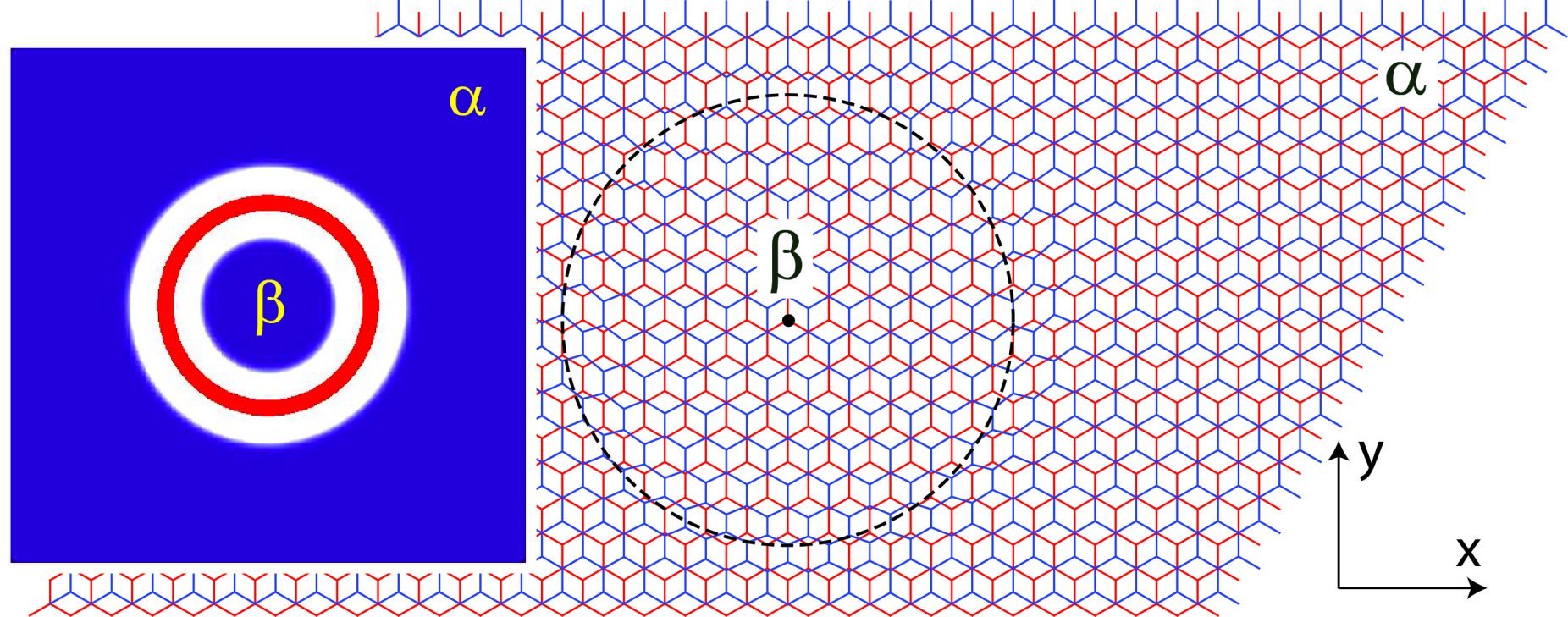}
  \caption{\label{sharpwall} A circular domain wall separating a disk of BLG with $\beta$ stacking from a background with $\alpha$
  stacking is produced by a rigid translation of its interior region by $\Delta_\beta - \Delta_\alpha$. This texture
  matches the interior and exterior textures through a domain wall (red) where every point on the wall is mapped to the state
  $s_r$ of Fig. 1.  The density plot in the inset gives the potential energy density and the wire-frame model of the lattice structure (background) illustrates
  the interlayer displacement field with the lattice constant greatly exaggerated for clarity.}
\end{figure}

Viewed on scales much larger than $\ell$ the field pattern in Fig. 2 changes sharply, and it is natural to consider other
textures that are globally smooth but maintain the same boundary conditions. Displacement fields with $\partial \Delta/\partial
z =0$ (antianalytic) are particularly useful since they are automatically divergenceless, avoiding any energy penalty due to
local compression or dilation, while minimizing its other nonzero strains. The smoothest such single-valued function that
satisfies the boundary conditions is
\begin{eqnarray}
\Delta^> = \Delta_\alpha + (\Delta_\beta - \Delta_\alpha) \bar z_0/(\bar z_0 - \bar z).
\end{eqnarray}
The left hand panel of Figure 3 illustrates this texture, where the density plot gives the potential energy density and the
lines show the folding of the symmetry lines of Fig. 1 onto the texture. In this mapping the sharp transition in the original
circular wall is collapsed to a single point $z_0$ represented by an isolated pole in $\Delta (\bar z)$. This singularity is
unavoidable for a nonconstant antianalytic field $\Delta^>$. In this mapping $\Delta(0) \equiv \Delta_\beta$ induces perfect
$\beta$ stacking at two additional points located at the same distance from its pole.

One can regard Eqn. 1 as describing an optimally smooth exterior elastic response to some, as yet unspecified, near field
displacement pattern. It can also be understood as the elastic response to an optimally smooth {\it interior} stacking pattern
in the region $|1-z/z_0|<1$ with
\begin{eqnarray}
\Delta^< = \Delta_\alpha + (\Delta_\beta - \Delta_\alpha) (z_0-z)/z_0
\end{eqnarray}
which is {\it analytic} in $z$ and matches $\Delta^>$ everywhere on a boundary as shown in Fig. 3 (right panel). Since
$\Delta^<$ is analytic it is automatically strain free, and since it is a linear analytic function of $z$ it has a constant
divergence ${\rm Re} [-(\Delta_\beta - \Delta_\alpha)/z_0]=0$. Physically this is required by the boundary conditions because
the interior solution has a {\it constant} compression/dilation which integrates to zero to match an exterior uncompressed
field $\Delta^>$ without any accumulation or depletion of material.

Since the modulus for long wavelength intralayer rotation is zero, the only elastic energy in the texture Eqns. (1-2) is the
exterior strain energy $U_e = \pi C |\Delta_\beta - \Delta_\alpha|^2/2 $ where $C$ is the two dimensional graphene shear
modulus ($\simeq 130 \, {\rm N/m}$) and is independent  of $z_0$  because of the scale invariance of the texture. This can be
compared with the elastic energy in a domain wall with line tension $\gamma$ which, for the geometry of Fig. 1, scales
extensively with the circumference $U_w = 2 \pi \gamma R$. Thus $U_w < U_e$ only for sufficiently small $R< K |\Delta_\beta -
\Delta_\alpha|^2 /(3 \gamma)$. For an estimated BLG line tension $\sim 100 \, {\rm pN}$ this requires $R< 7 \, {\rm nm}$ which
is essentially the domain wall width \cite{alden}. Thus, and as expected, the elastic energy generically favors spatially
varying solutions $\Delta^{>(<)}$ that are globally smooth.

\begin{widetext}

\begin{figure}
\includegraphics[angle=0,height=5cm]{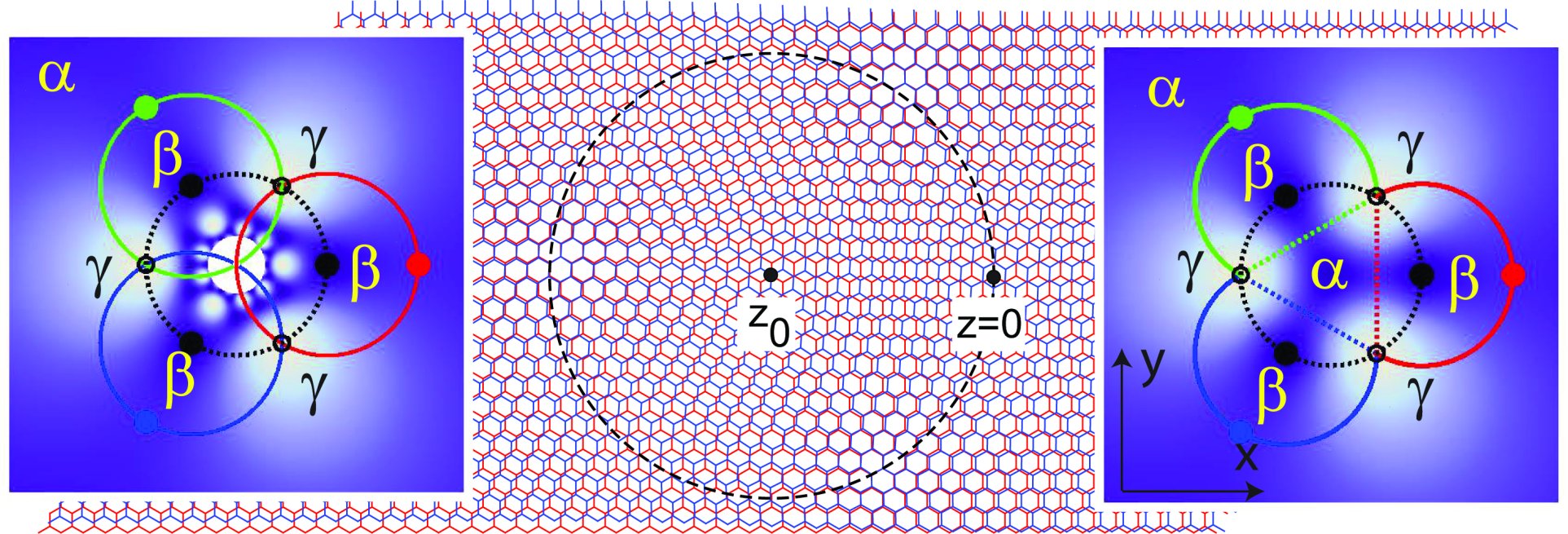}
  \caption{\label{smoothwall} Two globally smooth stacking textures with $\Delta(\infty) = \Delta_\alpha$ and $\Delta(0) =
  \Delta_\beta$. The density plots give the commensuration (potential) energy densities and the lines are the mapping
  of the color coded symmetry lines of Fig. 1 onto the textures. The left hand plot represents a texture that is antianalytic everywhere and contains
  a pole inside its core at $z=z_0$ (density plot is cut off at the white disk for clarity). The right hand plot represents a texture that matches an antianalytic function in the exterior
  region to an analytic function in the interior. The wire frame model of the lattice (background) illustrates the stacking pattern for the right
  hand texture (lattice constant greatly exaggerated for clarity.) }
\end{figure}

\end{widetext}

However these smooth solutions pay an energy penalty in their interlayer commensuration (potential) energies $U_c$. The
commensuration energy density $U(\Delta)$ (Fig. 1) increases quadratically near its extrema at $\Delta_{\alpha(\beta)}$:
$\delta U = \kappa_c |\Delta - \Delta_{\alpha(\beta)}|^2$ with $\kappa_c = 4 \pi^2 u_1/a^2$. In the far field the exterior
solution relaxes to its ground state algebraically with a logarithmically divergent energy $U_c = 8 \pi \kappa_c |\Delta_\beta
- \Delta_\alpha|^2  (|z_0|^2/\sqrt{3} a^2) \log R_c$ where $R_c$ is the system size. This excess field energy can be eliminated
by grouping $N$ defects in ``gauge neutral" clusters.  On scales large compared to the separation of these objects, with vector
charges $s_i$ located at positions $z_i$ $(i=1,N)$, the exterior displacement pattern is $ \Delta^> = \Delta_\alpha + \sum_i^N
s_i \bar z_i/(\bar z_i - \bar z)$ which eliminates the $1/\bar z$ tail when $\sum_i^N s_i \bar z_i=0$. An important case is
$N=3$ with $s_1=s_2=s_3$ and $z_2 = z_1 \exp(2 \pi i/3)$ and $z_3 = z_1 \exp(- 2\pi i/3)$ which breaks rotational symmetry by
the selection of a single direction $s_i$ for the triad, but nulls the monopole field by the threefold symmetry of the {\it
positions} of the defect centers. The left panel of Fig. 4 illustrates the field energy density and critical lines for one such
texture. The broken symmetry opens the ``red" boundary curve which then links the three defects, while the ``green" and ``blue"
critical lines form closed orbits that are confined around the individual defect centers.
\begin{figure}
  \includegraphics[angle=0,height=4.5cm]{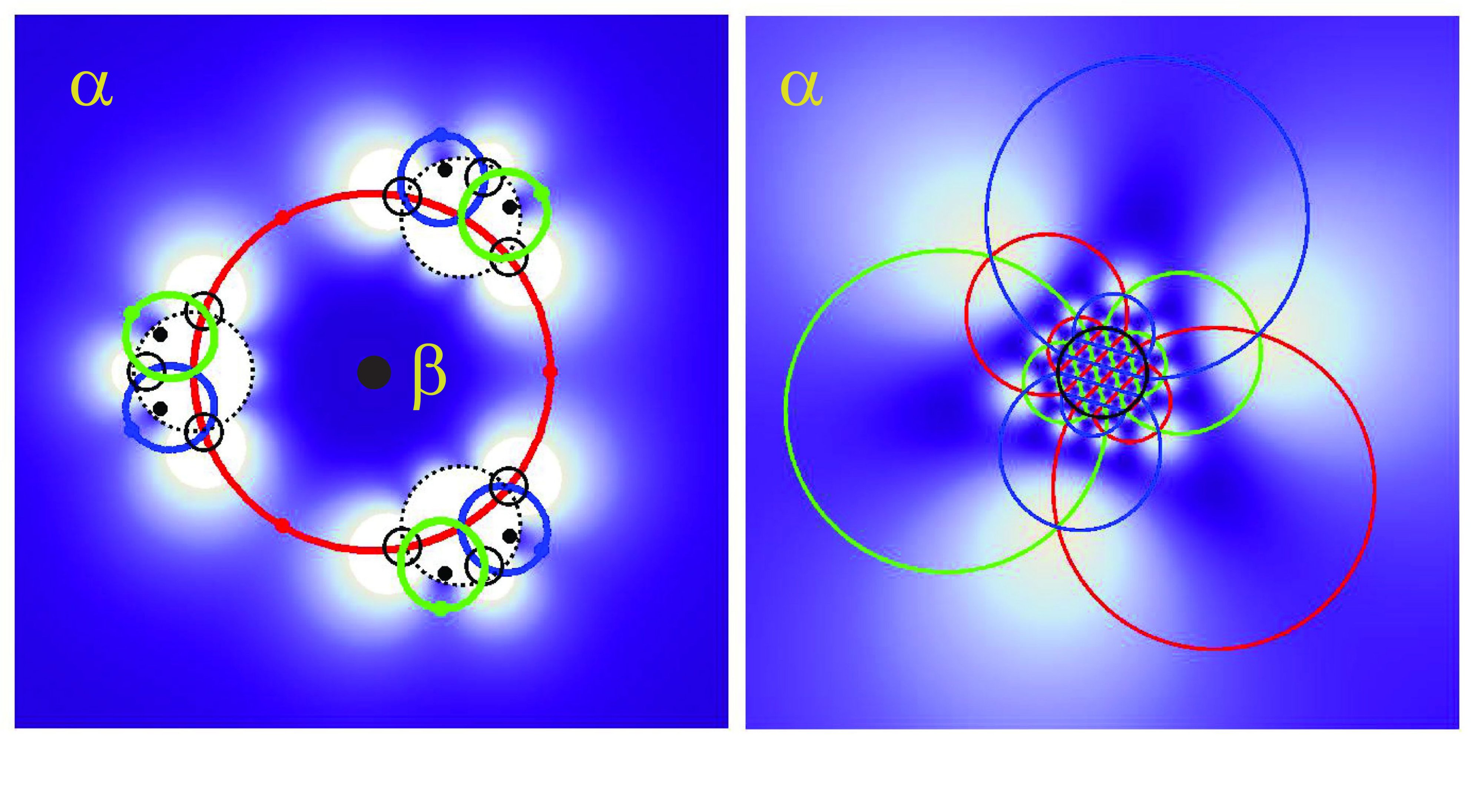}
  \caption{\label{cooltextures} (Left) A stacking texture that links three defects with equal strength $(\Delta_\beta -
  \Delta_\alpha)/3$ at three symmetry-related positions around the origin. The density plot gives the commensuration
  (potential) energy density and the lines map the color coded symmetry lines of Fig. 1 onto this texture. This
  combination of defects screens the long range tail of the texture and has a finite integrated commensuration energy. (Right)
  A stacking texture produced by clamping the origin in the $\beta$ stacked structure with $\Delta(0) = \Delta_\beta + T_1 +2T_2$,
  nucleating a region of twisted graphene in its interior, smoothly mapped to an incompressible and minimally
  strained exterior field.}
\end{figure}

One can iterate this process, uniformly distributing $N$ defects with charges $s_i$ on a circle at positions $R e^{i \phi_i}$
thereby cancelling its higher order multipoles.  In this case the texture has an expansion
\begin{eqnarray}
\Delta_N^< &=& \Delta_\alpha + \sum_i^N \, s_i \left( 1 - \frac{z e^{-i \phi_i}}{R} \right); \,\, |z|<R \nonumber\\
\Delta_N^> &=& \Delta_\alpha - \sum_{p\geq 1} \left(\frac{R}{\bar z} \right)^p \, \sum_i^N \, s_i e^{-i p \phi_i}; \,\, |z|>R
\end{eqnarray}
For $s_i = s_0 = (\Delta_\beta - \Delta_\alpha)/N$ and  $N \rightarrow \infty$ one obtains $\Delta^> = \Delta^\alpha$ and
$\Delta^< = \Delta_\beta$ representing a rigid interlayer translation for $|z|<R$ and reproducing the annular domain wall
pattern of Fig. 1. For general $N$, one can regard $\Delta_N$ with $s_i = s_0$ as a family of trial solutions where the value
of $N_{\rm min}$ is selected to minimize the sum of the elastic and commensuration energies $U_e +U_c$. $U_e(U_c)$ are
increasing(decreasing) functions of $N$, with a finite $N_{\rm min}$. The domain wall structures experimentally observed in BLG
are relatively wide \cite{alden,lin} indicating that the system is indeed in the regime dominated by the elastic energy,
favoring smooth (smaller $N$) over sharp (larger $N$) solutions. This is expected since the potential energy landscape has
quite broad minima in its low energy configuration space.

The textures given by Eqns. (1,2) realize identical $\alpha$ stackings at $|z| = \infty$ {\it and} in the near field at $z=z_0$
and are stable because of an additional boundary condition that clamps a different state at the origin: $\Delta(0) =
\Delta_\beta$. However, $\beta$ stacking at the origin occurs for a {\it lattice} of possible clamped states, each relatively
shifted by discrete lattice translation vectors $\Delta_{l,m}(0) = \Delta_\beta + l T_1 +m T_2$. Choices of $l$ and $m$ give
the winding of the order parameter $\Delta$ on the two cycles of a torus between $z=0$ and $z=\infty$ and index topologically
distinct solutions $\Delta_{l,m}$. Fig. 4 displays the field for the case $l=1$ and $m=2$. Remarkably, we find that the
interior solution represents a circular domain of uniformly rotated (twisted) graphene continuously matched to an untwisted and
minimally strained exterior texture. This illustrates a plausible mechanism for the formation of the observed complex stacking
textures in BLG. Isolated domains likely grow with uncorrelated local rotational registries forcing a complex stacking texture
in a state of minimum strain when the bilayer becomes continuous.

The textures identified here have well-studied analogs in (at least) two other physical contexts. First, they are similar,
though not identical to, the static baby skyrmion solutions of the two dimensional nonlinear sigma model \cite{volovik,piette}.
Normalizing our solution to its maximum value realized on the matching radius, $\Delta(z)/|\Delta(R)|= \vec n_\perp$ is the
$xy$ projection of a three dimensional unit vector $\hat n$. With the convention that the exterior(interior) regions map to the
upper(lower) hemisphere, our solution covers the sphere with degree $ (1/4 \pi) \int \, d^2 r \, \hat n \cdot \partial_1 \hat n
\times \partial_2 \hat n=-1$ and minimizes a projected strain energy functional $ U_{\rm BLG} = (1/2) \int \, d^2 r \, \nabla
\vec n_\perp \cdot \nabla \vec n_\perp$.  This breaks the full $O(3)$ symmetry of the nonlinear sigma model whose baby skyrmion
with the same degree instead minimizes $U_{{\rm nl}\sigma} = (1/2) \int \, d^2 r \, \nabla \hat n(\vec r) \cdot \nabla \hat
n(\vec r)$. The BLG solution is not obtained by a stereographic projection of the sphere onto the plane and has a slower far
field relaxation of its in plane components $\vec n_\perp$ and a faster relaxation of its (unmeasured) normal component $n_z$.
Second, our solutions are recognized as the classical field solutions from ordinary 2D electrostatics and magnetostatics where
they represent the field profile of a uniformly charged rod or  current carrying wire. An interesting difference is that in BLG
the field energy density vanishes for two nonzero values of the field (corresponding to degenerate vacua with $\alpha$ and
$\beta$ stacking) instead of just one state ($\vec E, \vec B =0)$.

Our approach provides a unified treatment of stacking point defects, domain walls and twisted graphene and provides a direction
for further investigation of these systems. For example it is possible that the complicated submicron structure observed in
these systems can be understood in terms of only a few fundamental stacking motifs and their conformal maps onto spatially
varying geometries imposed by pinning centers and irregularities in the sample morphology. More intriguingly, it is possible
that a desired BLG stacking texture could be controllably engineered using a combination of choice of substrate, growth face,
macroscopic curvature and various forms of submicron templating. Finally, although our model is designed to study static low
strain stacking configurations, they may also be important for nonlinear tribological properties of BLG, where defects of the
type studied here are generated when an applied mechanial load exceeds a critical yield stress \cite{hone2}.

This work was supported by the Department of Energy, Office of Basic Energy Sciences under Grant No. DE-FG02-ER45118. We thank
R. Kamien, C.L. Kane and F. Zhang for useful comments.

\newpage

\end{document}